\documentclass[]{article}

\usepackage{multirow}
\usepackage{amsmath}
\usepackage{amsfonts, amssymb}
\usepackage{amscd}

\newtheorem{lemm}{Lemma}
\newtheorem{theo}[lemm]{Theorem}
\newtheorem{prop}[lemm]{Proposition}
\newtheorem{defi}[lemm]{Definition}

\newcommand{\pr}{\indent{\em Proof:}}
\newcommand{\qed}{\hspace*{5 mm}$\Box$\bigskip}
\newenvironment{IEEEproof}{\noindent {\pr}}{\qed}

\newcommand{\wt}{\mbox{\rm w}_H}
\newcommand{\wl}{\mbox{\rm w}_L}
\newcommand{\Z}{{\mathbb{Z}}}

\newcommand{\dd}{\displaystyle}
\newcommand{\add}{\Z_2\Z_4}
\newcommand{\codi}{\mathcal{C}}

\newcommand{\zero}{{\mathbf{0}}}
\renewcommand{\u}{{\mathbf{1}}}
\renewcommand{\d}{{\mathbf{2}}}
\newcommand{\RM}{{\mathcal{RM}}}

\newcommand{\vv}{{\mathbf{v}}}
\newcommand{\vx}{{\mathbf{x}}}
\newcommand{\vy}{{\mathbf{y}}}
\newcommand{\vz}{{\mathbf{z}}}
\newcommand{\vu}{{\mathbf{u}}}
\newcommand{\vw}{{\mathbf{w}}}
\newcommand{\vhv}{{\mathbf{\hat{v}}}}
\newcommand{\vhu}{{\mathbf{\hat{u}}}}

\newcommand{\EQ}{\begin{equation}}
\newcommand{\EN}{\end{equation}}
\newcommand{\kip}[3]{\langle #1,#2 \rangle_{\otimes #3}}
\newcommand{\kipp}[2]{\langle #1,#2 \rangle_{\otimes N/4}}

\newcommand{\pcab}{\mathcal{PC}(\mathcal{A},\mathcal{B})}
\newcommand{\qpab}{\mathcal{QP}(\mathcal{A},\mathcal{B})}
\newcommand{\dpabcd}{\mathcal{DP}(\mathcal{A},\mathcal{B},\mathcal{C},\mathcal{D})}
\newcommand{\pc}{\mathcal{PC}}
\newcommand{\bqabc}{\mathcal{BQ}(\mathcal{A},\mathcal{B},\mathcal{C})}
\newcommand{\bqbcd}{\mathcal{BQ}(\mathcal{B},\mathcal{C},\mathcal{D})}
\newcommand{\bq}{\mathcal{BQ}}

\begin{document}

\title{Construction of $\Z_4$-linear Reed-Muller codes\thanks{This work has
been partially supported by the Spanish MEC and the European FEDER
Grant MTM2006-03250. Part of the material in Section III of this
paper was presented at the 17th Symposium on Applied algebra, Algebraic algorithms, and Error Correcting Codes (AAECC), Bangalore, India, December 2007.}}

\author{J.~Pujol, J.~Rif\`{a}, F.~I. Solov'eva
\thanks{J.~Pujol and J.~Rif\`{a} are with the Department of Information and
Communications Engineering, Universitat Aut\`{o}noma de Barcelona,
08193-Bellaterra, Spain.
 F.~I. Solov'eva is with the Sobolev Institute of Mathematics and Novosibirsk State
University, Novosibirsk, Russia.}}

\date{}
 \maketitle

\begin{abstract}
New quaternary Plotkin constructions are given and are used to
obtain new families of quaternary codes. The parameters of the
obtained codes, such as the length, the dimension and the minimum distance are studied. Using these constructions new families of
quaternary Reed-Muller codes are built with the peculiarity that
after using the Gray map the obtained $\Z_4$-linear codes have the
same parameters and fundamental properties as the codes in the usual binary linear
Reed-Muller family. To make more evident the duality relationships in the constructed families the concept of Kronecker inner product is introduced.
\end{abstract}


\section{Introduction}
In  \cite{N89} Nechaev   introduced the concept of
$\Z_4$-linearity of binary codes and later, in~\cite{Ham},
 Hammons, Kumar, Calderbank, Sloane and Sol\'e showed
that several families of binary codes are $\Z_4$-linear. In
\cite{Ham} it is proved that the binary linear Reed-Muller code
$RM(r,m)$ is $\Z_4$-linear for $r=0,1,2,m-1,m$ and is not
$\Z_4$-linear for $r=m-2$ ($m\geq 5$). In a subsequent work, Hou,
Lahtonen and Koponen~\cite{Hou}, proved that $RM(r,m)$ is
not $\Z_4$-linear for $3\leq r\leq m-2$.

 In \cite{Ham} it is introduced a construction of codes, called $\mathcal{QRM}(r,m)$, based on $\Z_4$-linear
 codes, such that after doing modulo two
 we obtain the usual binary linear Reed-Muller ($RM$) codes.
 In \cite{BFP05,BFP06} such family of codes is studied and their
 parameters are computed  as well as the dimension of the kernel and rank.
 In \cite{PR97} a kind of Plotkin construction was
used to build a  family of additive Reed-Muller codes and also in
\cite{Sol05} it was used a Plotkin construction to obtain a
sequence of quaternary linear Reed-Muller like codes. In both last
quoted constructions, the images of the obtained codes under the Gray
map are binary codes with the same parameters as the binary linear
$RM$ codes.  Moreover, on the other hand, in
\cite{BoRi99,Kro00,Kro01,prv}  were classified all the non-equivalent
$\add$-linear extended $1$-perfect codes and their duals, the
$\add$-linear Hadamard codes. It is a natural question to ask for
the existence of families of quaternary linear codes such that,
after the Gray map, the corresponding $\Z_4$-linear codes have the
same parameters as the well known family of binary linear $RM$
codes. In these new families, like in the usual $RM(r,m)$ family, the
code with $(r,m)=(1,m)$ should be a Hadamard code and the
code with $(r,m)=(m-2,m)$ should be an extended 1-perfect
code.

It is well known that an easy way to built the binary $RM$ family of
codes is by using the Plotkin construction \cite{Mac}. So,
it seems a good matter of study to try to generalize the Plotkin
construction to the quaternary linear codes and try to obtain new
families of codes which contain the above mentioned $\add$-linear
Hadamard codes and $\add$-linear extended $1$-perfect codes and
fulfill the same properties from a parameters point of view (length, dimension, minimum distance, inclusion and duality relationship) than
the binary $RM$ family.

In this paper we begin by studying the $\Z_4$-linear case and we organize it in the following way. In Section~\ref{sec:sec2} we introduce the concept of
quaternary code and give some constructions that could be seen as
quaternary generalizations of the well known binary Plotkin
construction. In Section~\ref{sec:sec3}, we construct several families
of $\Z_4$-linear Reed-Muller codes and  prove that they have similar
parameters as the classical binary $RM$ codes but they are not
linear. In Section~\ref{sec:sec4}, we discuss the concept of duality
for the constructed $\Z_4$-linear Reed-Muller codes and, finally, in
Section~\ref{sec:sec5} we give some conclusions and further research in
the topic.

\section{Constructions of quaternary codes}\label{sec:sec2}

\subsection{Quaternary codes}

Let $\Z_2$ and $\Z_4$ be the ring of integers modulo two and modulo four,
respectively. Let $\Z_2^n$ be the set of all binary vectors of length
$n$ and  $\Z_4^N$ be the set of all quaternary vectors of length
$N$. Any non-empty subset $C$ of $\Z_2^n$ is a
binary code and a subgroup of $\Z_2^n$ is called a {\it binary linear
code}. Equivalently, any non-empty
subset ${\cal C}$ of $\Z_4^N$ is a quaternary code and a subgroup of
$\Z_4^N$ is called a {\it quaternary linear code}. In general, any
non-empty subgroup ${\cal C}$ of $\Z_2^\alpha \times \Z_4^\beta$ is
an \textit{additive code}.

The Hamming weight $\wt(\vu)$ of a vector in $\Z_2^n$ is the
number of its nonzero coordinates. The Hamming distance
$d(\vu,\vv)$ between two vectors $\vu,\vv \in \Z_2^n$ is
$d(\vu,\vv)=\wt(\vu-\vv)$. For quaternary codes it is more
appropriate to use the Lee metric \cite{lee}. In $\Z_2$ the Lee
weight coincides with the Hamming weight, but in $\Z_4$ the Lee
weight of their elements is $\wl(0) = 0, \wl(1) = \wl(3) = 1$, and
$\wl(2) = 2$. The Lee weight $\wl(\vu)$ of a vector in $\Z_4^N$ is
the addition of the Lee weight of all the coordinates. The Lee
distance $d_L(\vu,\vv)$ between two vectors $\vu, \vv \in \Z_4^N$
is $d_L(\vu,\vv)= \wl(\vu-\vv)$.

Let $\codi$ be an \textit{additive code}, so a subgroup of
$\Z_2^{\alpha}\times\Z_4^{\beta}$ and let $C=\Phi(\codi)$, where
$\Phi: \Z_2^{\alpha}\times\Z_4^{\beta} \longrightarrow \Z_2^{n}$,
$n=\alpha+2\beta$, is given by $\Phi(\vu,\vv)=(\vu,\phi(\vv))$ for
any $\vu$ from $\Z_2^\alpha$ and any $\vv$ from $\Z_4^\beta,$
where $\phi:\Z_4^\beta\;\longrightarrow\;\Z_2^{2\beta}$ is the
usual Gray map, so $\phi(v_1,\ldots,v_\beta)=
(\varphi(v_1),\ldots,\varphi(v_\beta)),$ and
$\varphi(0)=(0,0),\varphi(1)=(0,1),\varphi(2)=(1,1)$,
$\varphi(3)=(1,0)$. We will use the symbols $\zero$, $\u$ and $\d$
for the all zeroes, the all ones and the all twos vectors,
respectively (by the context it will be always clear we speak
about the binary vectors $\zero$, $\u$ or quaternary, it will also
 be clear the length of the vectors).

Hamming and Lee weights, as well as Hamming and
Lee distances, can be generalized, in a natural way, to vectors
in $\Z_2^{\alpha}\times\Z_4^{\beta}$ by adding the corresponding
weights (or distances) of the  $\Z_2^\alpha$ part and the $\Z_4^\beta$ part.

Since $\codi$ is a subgroup of  $\Z_2^{\alpha}\times \Z_4^{\beta}$,
it is also isomorphic to an abelian structure like
$\Z_2^{\gamma}\times \Z_4^{\delta}$. Therefore, we have that
$|\codi|=2^\gamma 4^\delta $ and the number of order two codewords
in $\codi$ is $2^{\gamma+\delta}$. We call such code $\codi$ an \textit{additive code of type} $(\alpha,\beta;\gamma,\delta)$ and the binary
image $C=\Phi(\codi)$ a {\it $\Z_2\Z_4$-linear code of type
$(\alpha,\beta;\gamma,\delta)$}. In the specific case $\alpha=0$ we
see that $\codi$ is a quaternary linear code and its binary image is
called a {\it $\Z_4$-linear code}. Note that the binary length of
the binary code $C=\Phi(\codi)$ is $n=\alpha+2\beta$.

The minimum Hamming distance $d$ of a $\Z_2\Z_4$-linear code $C$
is the minimum value of $d(\vu,\vv)$, where $\vu,\vv \in C$ and
$\vu \neq \vv$. Notice that the Hamming distance of a
$\add$-linear code $C$ coincides with the Lee distance defined in
the additive code $\codi=\Phi^{-1}(C)$. From~now on, when we work
with distances it must be understood that we deal with
Hamming distances in the case of binary codes or Lee distances in
the case of additive codes.

Although $\codi$ could not have a basis, it is appropriate to
define a generator matrix for $\codi$ as
$$
   \mathcal{G}=\left ( \begin{array}{c|c}
        B_2 & Q_2 \\
        \hline
        B_1 & Q_1 \\
    \end{array}\right ),
$$
where $B_2$ is a $\gamma\times \alpha$ matrix; $Q_2$ is a
$\gamma \times \beta$ matrix; $B_1$ is a $\delta\times \alpha$
matrix and $Q_1$ is a $\delta\times \beta$ matrix. Matrices $B_1,
B_2$ are binary and $Q_1, Q_2$ are quaternary, but the entries in
$Q_2$ are only zeroes or twos.

Two additive codes $\codi_1$ and $\codi_2$ both of the same length
are said to be {\it monomially equivalent}, if one can be obtained
from the other by permuting the coordinates and multiplying by
$-1$ of certain coordinates. Additive codes which differ only by a
permutation of coordinates are said to be {\it permutation
equivalent}.

For $\Z_2\Z_4$-linear codes is usual to use the following definition
 of inner product in $\Z_2^{\alpha}\times
\Z_4^{\beta}$ that we will call the \emph{standard inner product}
\cite{RP97,BP07}:
 \EQ \label{eq:inner}
  \langle \vu,\vv \rangle=2(\sum_{i=1}^{\alpha} u_iv_i)+\sum_{j=\alpha+1}^{\alpha+\beta} u_jv_j\in \Z_4,
 \EN
where $\vu,\vv\in \Z_2^{\alpha}\times \Z_4^{\beta}$. We can also
write the standard inner product as
$$
    \langle \vu,\vv \rangle= \vu {\cdot} J_N{\cdot} \vv^{t},
$$
where $\dd J_N=\left (\begin{array}{c|c} 2I_{\alpha}& \zero \\
\hline \zero &I_{\beta}\end{array}\right )$, $N=\alpha+\beta$, is a
diagonal matrix over $\Z_4$. Note that when $\alpha=0$ the inner
product is the usual one for vectors over $\Z_4$ and when $\beta=0$
it is twice the usual one for vectors over $\Z_2$.

For $\alpha=0$ and $N=\beta=2^i$, $i=1,2,3,\ldots$, we can define
the inner product in an alternative way.
Let $K_2=\left (\begin{array}{cc} 1 & 0 \\
                0 & 3 \end{array} \right )$ be a matrix over $\Z_4$ and define
$K_N=\bigotimes_{j=1}^{\log_2(N)} K_2$ where $\bigotimes$ denotes
the Kronecker product of matrices. We call the \emph{Kronecker
inner product} the following: \EQ\label{kip} \langle \vu,\vv
\rangle_{\otimes N}= \vu {\cdot} K_N {\cdot} \vv^{t}. \EN

 The \textit{additive dual code} of $\codi$, denoted by ${\cal
C}^\perp$, is defined in the standard way as $${\cal
C}^\perp=\{\vu\in \Z_2^\alpha \times \Z_4^\beta \;|\;  \langle
\vu, \vv \rangle  =0 \mbox{ for all } \vv\in {\cal C}\}$$
or, using the Kronecker inner product
$${\cal
C}^\perp=\{\vu\in \Z_2^\alpha \times \Z_4^\beta \;|\;  \langle
\vu, \vv \rangle_{\otimes N}  =0 \mbox{ for all } \vv\in {\cal C}\}.$$

The definition and notations will be the same  for the
$\Z_4$-duality obtained by using the standard inner product or the Kronecker  inner product
and the difference will be
clear from the context.

Note that  $\langle \vu,\vv
\rangle_{\otimes N}= \vu {\cdot} K_N {\cdot} \vv^{t}=\langle \vu,\vv{\cdot}K_N
\rangle$. Hence, both additive dual codes by using the standard inner product or the Kronecker inner product, respectively, are monomially equivalent and so they have the same weight distribution. For both inner products, the additive dual code $\mathcal{C}^\perp$ is also an additive
code, that is a subgroup of $\Z_2^{\alpha}\times \Z_4^{\beta}$.
Its weight enumerator polynomial is related to the weight
enumerator polynomial of $\mathcal{C}$ by the MacWilliams identity
\cite{Del73}. The
corresponding binary code $\Phi({\cal C}^\perp)$ is denoted by
$C_\perp$ and called the {\it $\Z_2\Z_4$-dual code} of $C$. In the
case $\alpha=0$, the code ${\cal C}^\perp$ is also called the {\it
quaternary dual code} of ${\cal C}$ and $C_\perp$ the {\it
$\Z_4$-dual code} of $C$. Notice that $C$ and $C_\perp$ are not dual in the
binary linear sense but the weight enumerator polynomial of
$C_\perp$ is the McWilliams transform of the weight enumerator
polynomial of $C$. Given an additive code $\codi$ of type
$(\alpha,\beta,\gamma,\delta)$ it is  known  the type of the
additive dual code  (\cite{BP07} for additive codes with
$\alpha\not=0$ and \cite{Ham} for additive codes with $\alpha=0$).

In the present paper, as we will see later, the duality concept using the Kronecker inner product will make more visible the property that if a code $C$ belongs to a family of Reed-Muller codes then its dual code belongs to the same family.

From~now on, we focus our attention specifically to additive codes
with $\alpha=0$, so quaternary linear codes such that after the
Gray map they give rise to $\Z_4$-linear codes. Given a quaternary
linear code of type $(0,\beta;\gamma,\delta)$, we will write
$(N;\gamma,\delta)$ to say that $\alpha=0$ and $\beta=N$.

\subsection{The Plotkin construction}

In this section, we show that the well-known binary Plotkin construction can
be generalized to quaternary linear codes.

Let \, $\mathcal{A}$ \, and \, $\mathcal{B}$ \, be two  quaternary
linear codes of types \, $(N;\gamma_\mathcal{A},\delta_\mathcal{A})$
\, and $(N;\gamma_\mathcal{B},\delta_\mathcal{B})$ and minimum
distances $d_\mathcal{A}$, $d_\mathcal{B}$, respectively.
  Given $\vu\in \Z_4^{N}$ define $supp(\vu) \subset \{1,\ldots,N\}$ as  the set of nonzero coordinates of vector
 $\vu$.

\bigskip

\begin{defi}[Plotkin Construction]\label{defi:PlotDefi}
Given two quaternary linear codes $\mathcal{A}$ and $\mathcal{B}$,
we define a  quaternary linear code as
$$
    \pcab=\{(\vu_1|\vu_1+\vu_2):\vu_1\in \mathcal{A},\vu_2\in \mathcal{B}\}.
$$
\end{defi}
 It is easy to see that if
$\mathcal{G}_\mathcal{A}$ and $\mathcal{G}_\mathcal{B}$ are
generator matrices of $\mathcal{A}$ and $\mathcal{B}$, respectively, then the
matrix
$$
\mathcal{G}_{PC}=\left ( \begin{array}{cc}
                         \mathcal{G}_\mathcal{A} & \mathcal{G}_\mathcal{A} \\
                          0  & \mathcal{G}_\mathcal{B} \\
                      \end{array} \right )
$$
is a generator matrix of the  code $\pcab$.

\begin{prop}\label{prop:PlotConst}
The quaternary linear code $\pcab$ defined using the Plotkin construction is of type $(2N;\gamma, \delta)$, where
$\gamma=\gamma_\mathcal{A}+\gamma_\mathcal{B}$ and
$\delta=\delta_\mathcal{A}+\delta_\mathcal{B}$; the binary length is
$n=4N$; the size is $2^{\gamma+2\delta}$ and the minimum distance is
$d=\min\{2d_\mathcal{A},d_\mathcal{B}\}$.
\end{prop}

\begin{IEEEproof}
The type, the binary length and the size of $\pcab$ can be easily
computed from the definition of the code. The minimum distance can
be established as in the binary case \cite{Mac} but, by
completeness, we include the proof. Let us consider any vector
$\vu\in \pcab$ such that $\vu=(\vu_1|\vu_1+\vu_2)$, where $\vu_1
\in \mathcal{A}$
 and $\vu_2 \in \mathcal{B}$. Since $\pcab$ is a
 quaternary linear code, it is enough
 to prove that the weight $\wl(\vu)$ is not less than $d$.

 If $\vu_2=\zero$, then $\wl(\vu)=2\wl(\vu_1) \geq 2d_\mathcal{A}$.

 If $\vu_2\neq \zero$, by using the triangle inequality we immediately
 obtain
 $$
  \wl(\vu) = \wl(\vu_1)+\wl(\vu_1+\vu_2) \geq \wl(\vu_2)\geq  d_\mathcal{B}.
$$

 Hence $d \geq \min\{2d_\mathcal{A},d_\mathcal{B}\}$.
 The equality holds because taking the specific vectors
 $\vu_1 \in \mathcal{A}$ with minimum weight $d_\mathcal{A}$ and $\vu_2 \in \mathcal{B}$
with minimum weight $d_\mathcal{B}$ we obtain
$\wl(\vu_1|\vu_1)=2d_\mathcal{A}$
 and $\wl(\zero|\vv_2)= d_\mathcal{B}$.
  \end{IEEEproof}

\subsection{The quaternary Plotkin construction}

 A useful generalization of the above construction to obtain quaternary linear codes is the
following construction, called the {\it quaternary Plotkin
construction}. Such construction was used, for example, in
\cite{Kro01} for the classification of all $\Z_4$-linear Hadamard
codes.

\begin{defi}[Quaternary Plotkin Construction]\label{defi:QuatPlot}
Given two quaternary linear codes $\mathcal{A}$ and
$\mathcal{B}$, we define the quaternary linear code
$$
    \qpab=\{(\vu_1|\vu_1+\vu_2|\vu_1+2\vu_2|\vu_1+3\vu_2):\vu_1\in \mathcal{A},
    \vu_2\in \mathcal{B}\}.
$$
\end{defi}

It is easy to see that if $\mathcal{G}_\mathcal{A}$ and
$\mathcal{G}_\mathcal{B}$ are generator matrices of
$\mathcal{A}$ and $\mathcal{B}$, then the matrix
$$
\mathcal{G}_{QP}=\left ( \begin{array}{cccc}
                         \mathcal{G}_\mathcal{A} & \mathcal{G}_\mathcal{A} & \mathcal{G}_\mathcal{A} & \mathcal{G}_\mathcal{A}\\
                          0  & \mathcal{G}_\mathcal{B} & 2\mathcal{G}_\mathcal{B} & 3\mathcal{G}_\mathcal{B}\\
                      \end{array} \right )
$$
is a generator matrix of the  code $\qpab$.

\begin{prop}
The quaternary linear code $\qpab$ given in
Definition~\ref{defi:QuatPlot} is of  type
$(4N;\gamma, \delta)$, where
$\gamma=\gamma_\mathcal{A}+\gamma_\mathcal{B}$ and
$\delta=\delta_\mathcal{A}+\delta_\mathcal{B}$; the binary length is
$n=8N$; the size is $2^{\gamma+2\delta}$ and the minimum distance is $d\geq
\min\{4d_\mathcal{A},2d_\mathcal{B}\}$.
\end{prop}

\begin{IEEEproof}
The type, the binary length and the size of $\qpab$ can be easily
computed from the definition of the code. To check  the minimum
distance of $\qpab$ let us consider any vector
 $\vu\in \qpab$. Vector $\vu$ can be represented by
 $\vu=(\vu_1|\vu_1|\vu_1|\vu_1)+(\zero|\vu_2|2\vu_2|3\vu_2)$, where $\vu_1\in \mathcal{A}$
 and $\vu_2\in \mathcal{B}$. Since $\qpab$ is a quaternary linear code it is enough
  to show that the weight of $\vu$  is at least $d$.

  If $\vu_2=\zero$, then
$\wl(\vu)=4\wl(\vu_1) \geq 4d_\mathcal{A}$. The equality holds
taking a vector $\vu_1 \in \mathcal{A}$ of minimum weight.

For $\vu_2\neq \zero$ we have
\begin{equation*}
\begin{split}
\wl(\vu) & =\wl(\vu_1|\vu_1+\vu_2|\vu_1+2\vu_2|\vu_1+3\vu_2)\\
                           & = (\wl(\vu_1) +\wl(\vu_1+\vu_2))+(\wl(\vu_1+2\vu_2)+\wl(\vu_1+2\vu_2+\vu_2))\\
                           & \geq \wl(\vu_2)+ \wl(\vu_2)\;\mbox{(by using the triangle inequality)}\\
                           & \geq  2d_\mathcal{B}.
\end{split}
\end{equation*}
 \end{IEEEproof}

The Plotkin and the quaternary Plotkin constructions can be
combined in a \emph{double Plotkin construction}. Let \,
$\mathcal{A}$, \, $\mathcal{B}$, \, $\mathcal{C}$ and
$\mathcal{D}$ \, be four  quaternary linear codes of types \,
$(N;\gamma_\mathcal{A},\delta_\mathcal{A})$, \,
$(N;\gamma_\mathcal{B},\delta_\mathcal{B})$, \,
$(N;\gamma_\mathcal{C},\delta_\mathcal{C})$, \, and
$(N;\gamma_\mathcal{D},\delta_\mathcal{D})$  and minimum distances
$d_\mathcal{A}$, $d_\mathcal{B}$, $d_\mathcal{C}$,
$d_\mathcal{D}$, respectively.

\begin{defi}[Double Plotkin Construction]\label{defi:DoublePlot}
Given $\mathcal{A}$, $\mathcal{B}$,
$\mathcal{C}$ and $\mathcal{D}$ four quaternary linear codes, we define the
quaternary linear code
$$
\dpabcd=\{(\vu_1|\vu_1+\vu_2|\vu_1+2\vu_2+\vu_3|\vu_1+3\vu_2+\vu_3+\vu_4):\vu_1\in
\mathcal{A},\vu_2\in \mathcal{B},\vu_3\in \mathcal{C},\vu_4\in
\mathcal{D}\}.
$$
\end{defi}

It is easy to see that if $\mathcal{G}_\mathcal{A}$,
$\mathcal{G}_\mathcal{B}$, $\mathcal{G}_\mathcal{C}$ and
$\mathcal{G}_\mathcal{D}$ are generator matrices of
$\mathcal{A}$, $\mathcal{B}$, $\mathcal{C}$ and $\mathcal{D}$, then
the matrix
$$
\mathcal{G}_{DP}=\left ( \begin{array}{cccc}
                         \mathcal{G}_\mathcal{A} & \mathcal{G}_\mathcal{A} & \mathcal{G}_\mathcal{A} & \mathcal{G}_\mathcal{A}\\
                          0  & \mathcal{G}_\mathcal{B} & 2\mathcal{G}_\mathcal{B} & 3\mathcal{G}_\mathcal{B}\\
                          0  &              0          & \mathcal{G}_\mathcal{C} & \mathcal{G}_\mathcal{C}\\
                          0  &              0          &            0            & \mathcal{G}_\mathcal{D}\\
                      \end{array} \right )
$$
is a generator matrix of the  code $\dpabcd$.

\begin{prop}\label{prop:DoublePlot}
The quaternary linear code $\dpabcd$ given in
Definition~\ref{defi:DoublePlot} is of  type
$(4N;\gamma, \delta)$, where
$\gamma=\gamma_\mathcal{A}+\gamma_\mathcal{B}+\gamma_\mathcal{C}+\gamma_\mathcal{D}$ and
$\delta=\delta_\mathcal{A}+\delta_\mathcal{B}+\delta_\mathcal{C}+\delta_\mathcal{D}$; the binary length is $n=8N$; the size is $2^{\gamma+2\delta}$ and the minimum distance is
$d\geq
\min\{4d_\mathcal{A},2d_\mathcal{B},2d_\mathcal{C},d_\mathcal{D}\}$.
\end{prop}

\begin{IEEEproof}
The type, the binary length and the size of the code $\dpabcd$   can
be easily computed from the definition.

To check  the minimum distance of the code $\dpabcd$ let us
consider any vector $\vu$ from this code. It can be represented as
$\vu=
(\vu_1|\vu_1|\vu_1|\vu_1)+(\zero|\vu_2|2\vu_2|3\vu_2)+(\zero|\zero|\vu_3|\vu_3)+
(\zero|\zero|\zero|\vu_4)$, where $\vu_1 \in\mathcal{A}$,
$\vu_2\in \mathcal{B}$, $\vu_3\in \mathcal{C}$ and $\vu_4\in
\mathcal{D}$. Since $\dpabcd$ is a quaternary linear code it is
enough to show that the  weight of $\vu$ is, at least, $d$.

If $\vu_2= 0$ then we can write
$\vu=(\vu_1|\vu_1|\vu_1|\vu_1)+(\zero|\zero|\vu_3|\vu_3+\vu_4)$ so
that $\vu\in
\pc((\mathcal{A}|\mathcal{A}),\pc(\mathcal{C},\mathcal{D})),$
where $(\mathcal{A}|\mathcal{A})$ is the code generated by
$(\mathcal{G}_\mathcal{A}|\mathcal{G}_\mathcal{A})$. Using
Proposition \ref{prop:PlotConst} we obtain
$$\wl(\vu)=\min\{2d_{(\mathcal{A}|\mathcal{A})},d_{P(\mathcal{C},\mathcal{D})}\}=
\min\{4d_\mathcal{A},\min\{2d_\mathcal{C},d_\mathcal{D}\}\}=
\min\{4d_\mathcal{A},2d_\mathcal{C},d_\mathcal{D}\}.$$

 If $\vu_2\neq \zero$ then we distinguish two cases. If $\vu_4=\zero$ then
 $\wl(\vu)=\wl(\vu_1|\vu_1+\vu_2)+\wl(\vu_1+2\vu_2+\vu_3|\vu_1+3\vu_2+\vu_3)\geq
 \wl(\vu_2)+\wl(\vu_2)\geq 2d_{\mathcal{B}}$ using twice the triangle inequality.

 If $\vu_4\neq \zero$ then
 $\wl(\vu)=  \wl(\vu_1|\vu_1+\vu_2)+\wl(\vu_1+2\vu_2+\vu_3|\vu_1+3\vu_2+\vu_3+\vu_4)\geq
 \wl(\vu_2)+\wl(\vu_2+\vu_4)\geq \wl(\vu_4) \geq d_{\mathcal{D}}.$
 \end{IEEEproof}

Note that in case $\mathcal{B}=\mathcal{C}$ the bound is tight
because $d_\mathcal{B}=d_\mathcal{C}$ and the minimum distance
$d=\min\{4d_\mathcal{A},2d_\mathcal{C},d_\mathcal{D}\}$ can be
obtained taking specific vectors from $\mathcal{A}$, $\mathcal{C}$
or $\mathcal{D}$.

\subsection{The BQ-Plotkin construction}

We slightly change the construction given in Definition~\ref{defi:DoublePlot} in order to obtain a
tight bound for the minimum distance. We call this new construction the {\it BQ-Plotkin construction}.

Let $\mathcal{A}$, $\mathcal{B}$ and $\mathcal{C}$ be three
quaternary linear codes of types
$(N;\gamma_\mathcal{A},\delta_\mathcal{A})$,
$(N;\gamma_\mathcal{B},\delta_\mathcal{B})$,
$(N;\gamma_\mathcal{C},\delta_\mathcal{C})$, with minimum distances
$d_\mathcal{A}$, $d_\mathcal{B}$ and $d_\mathcal{C}$, respectively.

\begin{defi}[BQ-Plotkin Construction] \label{defi:BQPlot}
Let $\mathcal{G}_\mathcal{A}$, $\mathcal{G}_\mathcal{B}$ and
$\mathcal{G}_\mathcal{C}$ be generator matrices of the
quaternary linear codes $\mathcal{A}$, $\mathcal{B}$ and $\mathcal{C}$, respectively. We
define a new code $\bqabc$ as the quaternary linear code
generated by
$$
\mathcal{G}_{BQ} = \left ( \begin{array}{cccc}
                         \mathcal{G}_\mathcal{A} & \mathcal{G}_\mathcal{A} & \mathcal{G}_\mathcal{A} & \mathcal{G}_\mathcal{A} \\
                          0  & \mathcal{G}'_\mathcal{B} & 2\mathcal{G}'_\mathcal{B} & 3\mathcal{G}'_\mathcal{B} \\
                          0  &  0   & \hat{\mathcal{G}}_\mathcal{B} & \hat{\mathcal{G}}_\mathcal{B} \\
                          0  &  0   &  0  & \mathcal{G}_C \\
                      \end{array} \right ),
$$
where $\mathcal{G}'_\mathcal{B}$ is the matrix obtained from
$\mathcal{G}_\mathcal{B}$ after switching twos by ones in their
$\gamma_\mathcal{B}$  rows of order two and
$\hat{\mathcal{G}_\mathcal{B}}$ is the matrix obtained from
$\mathcal{G}_\mathcal{B}$ after removing their $\gamma_\mathcal{B}$
rows of order two.
\end{defi}

\begin{prop}\label{prop:BQPlotConst}
The quaternary linear code $\bqabc$ is  of  type
$(4N;\gamma,\delta)$, where
$\gamma=\gamma_\mathcal{A}+\gamma_\mathcal{C}$ and
$\delta=\delta_\mathcal{A}+\gamma_\mathcal{B}+2\delta_\mathcal{B}+\delta_\mathcal{C}$;
the binary length is $n=8N$; the size is $2^{\gamma+2\delta}$ and
the minimum distance $d =
\min\{4d_\mathcal{A},2d_\mathcal{B},d_\mathcal{C}\}$.
\end{prop}

\begin{IEEEproof}
The type, the length and the size of  $\bqabc$ can be easily computed from
the definition of the code.

To check  the minimum distance of  $\bqabc$ let us consider
any vector
 $\vu=(\vu_1|\vu_1|\vu_1|\vu_1)+(\zero|\vu_2|2\vu_2|3\vu_2)+(\zero|\zero|\vu_3|\vu_3)+
 (\zero|\zero|\zero|\vu_4)\in \bqabc$,
 where $\vu_1\in \mathcal{A}$; $\vu_2\in \mathcal{B}'$;
 $\vu_3\in \hat{\mathcal{B}}$ and $\vu_4\in \mathcal{C}$. Codes $\mathcal{B}'$ and
 $\hat{\mathcal{B}}$ are
the quaternary linear codes generated by
$\mathcal{G}'_\mathcal{B}$ and $\hat{\mathcal{G}}_\mathcal{B}$,
respectively. Since $\bqabc$ is a quaternary linear code it is
enough
  to show that the weight of $\vu$  is at least $d$.

If $\vu_2=\zero$ then by using the same arguments as in
Proposition \ref{prop:DoublePlot} we have $\wl(\vu)\geq
\min\{4d_{\mathcal{A}},2d_{\mathcal{B}},d_{\mathcal{C}}\}$ because
$d_{\hat{\mathcal{B}}} \geq d_{\mathcal{B}}$.

If $\vu_2\neq \zero$ then we distinguish two cases. If
$\vu_4=\zero$ then
\begin{equation*}
\begin{split}
 \wl(\vu) & =\wl(\vu_1|\vu_1+\vu_2|\vu_1+2\vu_2+\vu_3|\vu_1+3\vu_2+\vu_3)\\
 &= (\wl(\vu_1) +\wl(\vu_1+\vu_2))+(\wl(\vu_1+2\vu_2+\vu_3)+\wl(\vu_1+2\vu_2+\vu_3+\vu_2))\\
 &= (\wl(\vu_1) + \wl(\vu_1+2\vu_2+\vu_3)) + (\wl(\vu_1+\vu_2))+\wl(\vu_1+2\vu_2+\vu_3+\vu_2))\\
 & \geq \wl(2\vu_2+\vu_3)+ \wl(2\vu_2+\vu_3)\;\mbox{(by the triangle inequality)}\\
 & \geq  2\wl(2\vu_2+\vu_3).
\end{split}
\end{equation*}

Note that $2\vu_2\in \mathcal{B}$ and $\vu_3 \in \hat{\mathcal{B}}
\subset \mathcal{B}$. If $\vu_3\neq 2\vu_2$ then
$\wl(2\vu_2+\vu_3)\geq d_\mathcal{B}$ and $\wl(\vu) \geq
2d_\mathcal{B}$. If $2\vu_2=\vu_3$ then $\vu_2 \in
\hat{\mathcal{B}}$ and $\wl(\vu_2)\geq d_\mathcal{B}$. So,
$\wl(\vu) = \wl(\vu_1|\vu_1+\vu_2|\vu_1|\vu_1+\vu_2)\geq
2\wl(\vu_2)\geq 2d_\mathcal{B}$.

Using twice the triangle inequality, the case $\vu_4\neq 0$ easily
gives
\begin{equation*}
\begin{split}
\wl(\vu) & = \wl(\vu_1|\vu_1+\vu_2) + \wl(\vu_1+2\vu_2+\vu_3|\vu_1+3\vu_2+\vu_3+\vu_4)\\
                                                 & \geq \wl(\vu_2) + \wl(\vu_2+\vu_4) \geq  \wl(\vu_4).
\end{split}
\end{equation*}

Hence, $d \geq \min\{4d_\mathcal{A},2d_\mathcal{B},d_\mathcal{C}\}$. But the
equality holds after the following considerations.

Taking the specific vector $\vu_1 \in \mathcal{A}$ with minimum
weight $d_\mathcal{A}$ we obtain
$\wl(\vu_1|\vu_1|\vu_1|\vu_1)=4d_\mathcal{A}$.

Taking the specific vector $\vu_4 \in \mathcal{C}$ with minimum
weight $d_\mathcal{C}$ we obtain
$\wl(\zero|\zero|\zero|\vu_4)=d_\mathcal{C}$.

Taking the specific vector $\vu_2 \in \mathcal{B}$ with minimum
 weight $d_\mathcal{B}$ we obtain the following.  Note that $\mathcal{B} \subset \mathcal{B}'$ and so we can write the vector $\vu_2$ as $\vu_2 = \vhv+2\vw'$, where $\vhv \in \hat{\mathcal{B}}$ and $\vw' \in \mathcal{B}'\backslash \hat{\mathcal{B}}$.  Take the vector $\vu_2=\vhv+2\vw'\in \mathcal{B}'$ and, moreover, the vector $\vhu=2\vhv\in \hat{\mathcal{B}}$ and compose the vector $(0|\vu_2|2\vu_2+\vhu|3\vu_2+\vhu)=(0|\vu_2|0|\vu_2)$ which belongs to  $\bqabc$. This vector has minimum Lee weight $2d_\mathcal{B}$.
 \end{IEEEproof}

\section{Quaternary Reed-Muller codes}\label{sec:sec3}

The usual binary linear $RM$ family of codes is one of the oldest
 and interesting family of codes. The codes in this family are easy
 to decode and their combinatorial properties are of great interest to
 produce new optimal codes.

For any integer $m\geq 1$ the family of binary linear $RM$ codes is
given by the sequence $RM(r,m)$, where $0\leq r \leq m$. The code
$RM(r,m)$ is called the $r$-th order binary linear Reed-Muller code
of length $n=2^m$ and it is true that
$$
    RM(0,m) \subset RM(1,m) \subset \ldots \subset RM(r-2,m)\subset
    RM(r-1,m)\subset RM(r,m).
$$

Let $0\leq r\leq m$, $m\geq 1$. Following~\cite{Mac} the $RM(r,m)$
code of order $r$
 can be constructed by using the Plotkin construction in the following
way:
\begin{eqnarray}\label{defi:definicio}
     RM(0,m)&= & \{\zero,\u\}, \,\,\,  RM(m,m)\,=\,\Z_2^{2^m},\nonumber\\
     RM(r,m)& = & \{(\vu_1|\vu_1+\vu_2):\vu_1\in RM(r,m-1),\, \vu_2\in RM(r-1,m-1)\}.
\end{eqnarray}

It is important to note that if we fix $m$, once we know the
sequence $RM(r,m)$ for all $0\leq r\leq m$, then it is easy to obtain
the new sequence $RM(r,m+1)$ by using the Plotkin construction
(\ref{defi:definicio}).

Moreover, the codes in the $RM$ family fulfill the  basic properties
summarized in the following theorem:

\begin{theo}[\cite{Mac}]\label{theo:reed-muller}
The binary linear Reed-Muller family of codes $\{RM(r,m)\}$, $0\leq r \leq m$, has the following
 properties:
\begin{enumerate}
    \item the length $n=2^m$, $m\geq 1$;
    \item the minimum distance $d=2^{m-r}$;
    \item the dimension $\dd k=\sum_{i=0}^r \binom{m}{i}$;
    \item the code $RM(r-1,m)$ is a subcode of $RM(r,m)$, $r>0$.
    The code $RM(0,m)$ is the repetition code with only one
        nonzero codeword (the all ones vector). The code
        $RM(m,m)$ is the whole space $\Z_2^{2^m}$ and
          $RM(m-1,m)$ is the even code (that is, the code with
          all the vectors of even weight from $\Z_2^{2^m}$);
    \item the code $RM(1,m)$ is the binary linear Hadamard code and $RM(m-2,m)$ is the extended
          binary Hamming code of parameters $(2^m, 2^m-m-1,4)$;
    \item the code $RM(r,m)$  is the dual code of $RM(m-1-r,m)$ for $0\leq r<m$.
\end{enumerate}
\end{theo}

In the recent literature \cite{Ham,W97,BFP05,BFP06} several families
of quaternary linear codes have been proposed and studied  trying to
generalize the $RM$ codes. However, when we take the corresponding
$\Z_4$-linear codes, they do not satisfy all the properties in
Theorem~\ref{theo:reed-muller}. This last requirement is the main
goal of the present work, to construct new families of quaternary
linear codes such that, after the Gray map, we obtain $\Z_4$-linear
codes with the parameters and properties quoted in
Theorem~\ref{theo:reed-muller}. The result of the present paper
generalizes the results in~\cite{Sol05}.

\medskip

Further we will refer to these quaternary linear Reed-Muller codes
as $\RM$ to distinguish them from the binary linear Reed-Muller
codes $RM$. Contrary to the binary linear case, where there is only
one $RM$ family, in the quaternary case we have
$\lfloor\frac{m+1}{2}\rfloor$ families for each value of $m$. We
will distinguish these families by using subindexes $s$ from the set
$\{0,\ldots,\lfloor\frac{m-1}{2}\rfloor\}$.

\subsection{The family of $\RM(r,1)$ codes}\label{sec:ARM2}

We begin by considering the trivial case of $m=1$, that is, the case
of codes of binary length $n=2^1$. The quaternary linear Reed-Muller
code $\RM(0,1)$ is the repetition code with only one nonzero
codeword (the vector with only one quaternary coordinate of value
$2$). This quaternary linear code is of type $(1;1,0)$. The code
$\RM(1,1)$   is the whole space $\Z_4^1$, so a quaternary linear
code of type $(1;0,1)$.

These two codes, $\RM(0,1)$ and $\RM(1,1)$, after the Gray map,
give binary codes with the same parameters of the corresponding
binary codes $RM(r,1)$  and with the same properties described in
Theorem~\ref{theo:reed-muller}. In this case, when $m=1$, not only
these codes have the same parameters, but they have the same
codewords.

We will refer to these codes as $\RM_0(0,1)$ and $\RM_0(1,1)$,
respectively, as it is shown in Table~\ref{tab:taulaARM(r,1)}. In
each entry of this table there are the parameters $(\gamma,\delta)$
of the corresponding code of type ($N;\gamma,\delta$).

\begin{table}
\caption{$\RM(r,m)$ codes for $m=1$}\label{tab:taulaARM(r,1)}
\begin{center}
    \begin{tabular}{c|cc|c}
            \cline{2-3} & \multicolumn{2}{c|}{$(r,m)$} & \\
            \cline{2-3} & (0,1)&(1,1) & \\
            \cline{2-3} $N$ & \multicolumn{2}{c|}{($\gamma$,$\delta$)}& \\
            \hline\hline
            1 &(1,0)&(0,1)& $\RM_0(r,1)$\\
            \cline{2-3}
    \end{tabular}
\end{center}
\end{table}

Since we will need an specific representation for these codes in
Table~\ref{tab:taulaARM(r,1)}, we will agree in using further the
following matrices as the generator matrices for each one of
them. The generator matrix of $\RM_0(0,1)$ is $\mathcal{G}_0(0,1)=\left(\begin{array}{c}2\\
\end{array}\right)$ and the generator matrix of $\RM_0(1,1)$ is
$\mathcal{G}_0(1,1)=\left(\begin{array}{c}1\\
\end{array}\right)$.

\subsection{Plotkin and BQ-Plotkin constructions}\label{sec:construccions}

The first important point is to apply the Plotkin construction to
quaternary linear Reed-Muller codes.

Let $\RM_s(r,m-1)$ and $\RM_s(r-1,m-1)$,
 $0\leq s\leq \lfloor\frac{m-2}{2}\rfloor $, be any two $\RM$ codes
of type ($N;\gamma_{r,m-1}^s,\delta_{r,m-1}^s$) and
($N;\gamma_{r-1,m-1}^s,\delta_{r-1,m-1}^s$); binary length
$n=2^{m-1}$; number of codewords $2^{k_r}$ and $2^{k_{r-1}}$;
minimum distance $2^{m-r-1}$ and $2^{m-r}$ respectively, where
$$
  k_r= \dd \sum_{i=0}^r\binom{m-1}{i},\,\,\,\, k_{r-1}=\dd \sum_{i=0}^{r-1}\binom{m-1}{i}.
$$

\begin{theo}\label{theo:plotkin}
For any $r$ and $m\geq 2$, $0< r< m$, the code obtained by using
the Plotkin construction
$$
    \RM_s(r,m)=\{(\vu_1|\vu_1+\vu_2):\vu_1\in \RM_s(r,m-1),\, \vu_2\in \RM_s(r-1,m-1)\},
$$
where $0\leq s\leq \lfloor\frac{m-1}{2}\rfloor,$ is a quaternary
linear code of type ($2N$; $\gamma_{r,m}^s$, $\delta_{r,m}^s$),
where $\gamma_{r,m}^s=\gamma_{r,m-1}^s+\gamma_{r-1,m-1}^s$ and
$\delta_{r,m}^s=\delta_{r,m-1}^s+\delta_{r-1,m-1}^s$; the binary
length is $n=2^{m}$; the number of codewords is $2^k$, where $k= \dd
\sum_{i=0}^r\binom{m}{i}$, the code distance is $2^{m-r}$ and
$\RM_s(r-1,m) \subset \RM_s(r,m)$.

For $r=0$, the code $\RM_s(0,m)$ is the repetition code with only
one nonzero codeword (the all twos vector). For $r=m$, the code
$\RM_s(m,m)$  is the whole space $\Z_4^{2^{m-1}}$.
\end{theo}

\begin{IEEEproof}
The type  ($2N; \gamma_{r,m}^s, \delta_{r,m}^s$) of the
 code $\RM_s(r,m)$, its size and the minimum distance can be computed
from Proposition~\ref{prop:PlotConst}. Since $\RM_s(r-1,m-1) \subset
\RM_s(r,m-1)$ and $\RM_s(r-2,m-1) \subset \RM_s(r-1,m-1)$, then
taking into account the codes given in the previous section by
induction we get $\RM_s(r-1,m) \subset \RM_s(r,m)$.
 \end{IEEEproof}

\bigskip

For $m=2$, taking the $\RM_0(r,1)$ codes in
Table~\ref{tab:taulaARM(r,1)} and applying
Theorem~\ref{theo:plotkin} we obtain the codes in
Table~\ref{tab:taulaARM(r,2)}. The generator matrices for these
codes are the following \EQ\label{eq:m=2}
    \RM_0(0,2):\left(\begin{array}{cc}2&2\\\hline
\end{array}\right);\,\,\RM_0(1,2):\left(\begin{array}{cc}0&2\\\hline
1&1\end{array}\right); \RM_0(2,2):\left(\begin{array}{cc}\hline
1&0\\0&1\end{array}\right).
\EN

\begin{table}\caption{$\RM(r,m)$ codes for $m=2$}\label{tab:taulaARM(r,2)}
\begin{center}$\begin{array}{c||ccc|c}
        \cline{2-4} & \multicolumn{3}{c|}{(r,m)}& \\
        \cline{2-4} & (0,2)&(1,2)&(2,2) & \\
        \cline{2-4} N & \multicolumn{3}{c|}{(\gamma,\delta)}& \\
        \hline\hline
        2 &(1,0)&(1,1)&(0,2)&\RM_0(r,2)\\
        \cline{2-4}
\end{array}$
\end{center}
\end{table}

For $m=3$, it is well known that there exist two $\Z_4$-linear
Hadamard codes \cite{Kro01}. So, our goal is to construct two
families of quaternary linear Reed-Muller codes as it is shown in
Table~\ref{tab:taulaARM(r,3)}. The codes in the first row of
Table~\ref{tab:taulaARM(r,3)} can be obtained due to the Plotkin
construction from the codes of Table~\ref{tab:taulaARM(r,2)}. But,
the codes in the second row can not be obtained by using only the
Plotkin construction. It is in this case that we need to exploit
the new BQ-Plotkin construction as we will see later in this
section.

\begin{table}\caption{$\RM(r,m)$ codes for $m=3$}\label{tab:taulaARM(r,3)}
\begin{center}
$\begin{array}{c||cccc|c}
    \cline{2-5} & \multicolumn{4}{c|}{(r,m)} & \\
    \cline{2-5} & (0,3)&(1,3)&(2,3)&(3,3) &\\
    \cline{2-5} N & \multicolumn{4}{c|}{(\gamma,\delta)}& \\
    \hline\hline
    4 &(1,0)&(2,1)&(1,3)&(0,4)&\RM_0(r,3)\\
    4 &(1,0)&(0,2)&(1,3)&(0,4)&\RM_1(r,3)\\
    \cline{2-5}
\end{array}$
\end{center}
\end{table}

Constructions of additive  codes with the parameters of the binary
linear Reed-Muller  codes by using only the Plotkin construction
were initiated in \cite{PR97,Sol05}.

\bigskip

\begin{lemm}\label{lemm:PBQ-subset}
Let $\{\mathcal{A}_i\}$, $i=1,2,3,4$, be a family of four
quaternary linear codes of types $(N;\gamma_i,\delta_i)$ with
generator matrices $\mathcal{G}_i$, respectively. Let
$\mathcal{A}_i'$ and $\hat{\mathcal{A}}_i$ be the codes generated
by $\mathcal{G}_i'$ and $\hat{\mathcal{G}}_i$, respectively, such
that for $i=1,2,3$ it is true that

\begin{description}
 \item[(i)] $\mathcal{A}_i\subset \mathcal{A}_{i+1}$;
 \item[(ii)] $\hat{\mathcal{A}}_i\subset \hat{\mathcal{A}}_{i+1}$;
 \item[(iii)] $\mathcal{A}_i'\subset \mathcal{A}_{i+1}'$;
 \item[(iv)] $\mathcal{A}_i'\subset \mathcal{A}_{i+1}$.
 \end{description}
Then, the family $\{\pc(\mathcal{A}_{i+1},\mathcal{A}_i)\}$ of the
three codes $\pc(\mathcal{A}_2,\mathcal{A}_{1})$,
$\pc(\mathcal{A}_3,\mathcal{A}_{2})$ and
$\pc(\mathcal{A}_4,\mathcal{A}_{3})$ satisfies  \textit{(i)},
\textit{(ii)}, \textit{(iii)} and \textit{(iv)} for $i=1,2$ and the
family $\{ \bq(\mathcal{A}_{i+2},\mathcal{A}_{i+1},\mathcal{A}_i)\}$
of the two codes
$\bq(\mathcal{A}_3,\mathcal{A}_{2},\mathcal{A}_{1})$ and
$\bq(\mathcal{A}_4,\mathcal{A}_{3},\mathcal{A}_{2})$ satisfies the
properties \textit{(i)}, \textit{(ii)}, \textit{(iii)} and
\textit{(iv)} for $i=1$.
\end{lemm}

\begin{IEEEproof}
It is straightforward to see that the Plotkin construction fulfills the properties.

For the BQ-Plotkin construction the property \emph{(i)} is clear
from Definition~\ref{defi:BQPlot}.

Now, the generator matrix of
$\hat{\bq}(\mathcal{A}_{i+2},\mathcal{A}_{i+1},\mathcal{A}_{i})$ has
the following form:
 \EQ \label{eq:BQhat}
 \left ( \begin{array}{cccc}
      \hat{\mathcal{G}}_{i+2} & \hat{\mathcal{G}}_{i+2} & \hat{\mathcal{G}}_{i+2} & \hat{\mathcal{G}}_{i+2} \\
      0  & \mathcal{G}_{i+1}' & 2\mathcal{G}_{i+1}' & 3\mathcal{G}_{i+1}' \\
      0  &  0   & \hat{\mathcal{G}}_{i+1} & \hat{\mathcal{G}}_{i+1} \\
      0  &  0   &  0  & \hat{\mathcal{G}_i} \\
    \end{array} \right ).
 \EN
Using the properties \emph{(ii)} and \emph{(iii)} for the matrices
\, $\mathcal{G}_{i},$ \, $\mathcal{G}_{i+1},$ \,
$\mathcal{G}_{i+2}$, \, we get the property \emph{(ii)} for
$\bq(\mathcal{A}_{i+2},\mathcal{A}_{i+1},\mathcal{A}_{i})$ and
$i=1$.

Since, the generator matrix of
$\bq'(\mathcal{A}_{i+2},\mathcal{A}_{i+1},\mathcal{A}_{i})$ has the
form
 \EQ \label{eq:BQprime}
    \left ( \begin{array}{cccc}
      \mathcal{G}_{i+2}' & \mathcal{G}_{i+2}' & \mathcal{G}_{i+2}' & \mathcal{G}_{i+2}' \\
      0  & \mathcal{G}_{i+1}' & 2\mathcal{G}_{i+1}' & 3\mathcal{G}_{i+1}' \\
      0  &  0   & \hat{\mathcal{G}}_{i+1} & \hat{\mathcal{G}}_{i+1} \\
      0  &  0   &  0  & \mathcal{G}_i' \\
    \end{array} \right ),
 \EN
using the properties \emph{(ii)} and \emph{(iii)} for the matrices
$\mathcal{G}_{i}, \mathcal{G}_{i+1}, \mathcal{G}_{i+2}$ we obtain
the property \emph{(iii)} for
$\bq(\mathcal{A}_{i+2},\mathcal{A}_{i+1},\mathcal{A}_{i})$ and
$i=1$.

Finally, the generator matrix of
$\bq(\mathcal{A}_{i+2},\mathcal{A}_{i+1},\mathcal{A}_{i})$ has the
form
 \EQ \label{eq:BQstandard}
    \left ( \begin{array}{cccc}
      \mathcal{G}_{i+2} & \mathcal{G}_{i+2} & \mathcal{G}_{i+2} & \mathcal{G}_{i+2} \\
      0  & \mathcal{G}_{i+1}' & 2\mathcal{G}_{i+1}' & 3\mathcal{G}_{i+1}' \\
      0  &  0   & \hat{\mathcal{G}}_{i+1} & \hat{\mathcal{G}}_{i+1} \\
      0  &  0   &  0  & \mathcal{G}_i \\
    \end{array} \right ).
 \EN
Using the properties \emph{(ii)}, \emph{(iii)} and \emph{(iv)} for
the matrices $\mathcal{G}_{i}, \mathcal{G}_{i+1},
\mathcal{G}_{i+2}$ we obtain the property  \emph{(iv)} for
$\bq(\mathcal{A}_{i+2},\mathcal{A}_{i+1},\mathcal{A}_{i})$ and
$i=1$.
\end{IEEEproof}

\bigskip

Let $\RM_{s-1}(r,m-2)$, $\RM_{s-1}(r-1,m-2)$ and
$\RM_{s-1}(r-2,m-2)$,
  $0< s\leq \lfloor\frac{m-3}{2}\rfloor $, $m> 3$, be any three $\RM$ codes
of type \, ($N;\gamma_{r,m-2}^{s-1},\delta_{r,m-2}^{s-1}$), \,
($N;\gamma_{r-1,m-2}^{s-1},\delta_{r-1,m-2}^{s-1}$) and
($N;\gamma_{r-2,m-2}^{s-1},\delta_{r-2,m-2}^{s-1}$); binary
length $n=2^{m-2}$; number of codewords $2^{k_r}$,
$2^{k_{r-1}}$ and $2^{k_{r-2}}$; minimum distances
$2^{m-r-2}$, $2^{m-r-1}$ and $2^{m-r}$ respectively, where
$$
 k_r= \dd \sum_{i=0}^r\binom{m-2}{i},\,\,\,\, k_{r-1}=\dd \sum_{i=0}^{r-1}\binom{m-2}{i},
 \,\,\,\, k_{r-2}=\dd \sum_{i=0}^{r-2}\binom{m-2}{i}.
$$
Let $\mathcal{G}_{s}(r,m)$, $0<r<m-1$, be the matrix
\EQ\label{matrix_even}\left (
\begin{array}{cccc}
                         \mathcal{G}_{s-1}(r,m-2) & \mathcal{G}_{s-1}(r,m-2) & \mathcal{G}_{s-1}(r,m-2) & \mathcal{G}_{s-1}(r,m-2) \\
                          0  & \mathcal{G}_{s-1}'(r-1,m-2) & 2\mathcal{G}_{s-1}'(r-1,m-2) & 3\mathcal{G}_{s-1}'(r-1,m-2) \\
                          0  &  0   & \hat{\mathcal{G}}_{s-1}(r-1,m-2) & \hat{\mathcal{G}}_{s-1}(r-1,m-2) \\
                          0  &  0   &  0  & \mathcal{G}_{s-1}(r-2,m-2) \\
                      \end{array} \right )
\EN

For the special case $r=1$ we need to define
$\mathcal{G}_{s-1}(-1,m-2)$ as the generator matrix of the all zero
codeword code.

\bigskip

\begin{theo}\label{theo:BQ-plotkin}
For any $r$ and $m\geq 3$, $0< r< m-1$, the  $\RM_s(r,m)$ code, $0
< s\leq \lfloor\frac{m-1}{2}\rfloor $, obtained by using the
BQ-Plotkin construction in Definition~\ref{defi:BQPlot} and with the
generator matrix $\mathcal{G}_s(r,m)$ defined in
(\ref{matrix_even}), is a quaternary linear code of type ($4N$;
$\gamma_{r,m}^s$, $\delta_{r,m}^s$), where
$\gamma_{r,m}^s=\gamma_{r,m-2}^{s-1}+\gamma_{r-2,m-2}^{s-1}$;
$\delta_{r,m}^s=\delta_{r,m-2}^{s-1}+\gamma_{r-1,m-2}^{s-1}+
2\delta_{r-1,m-2}^{s-1}+\delta_{r-2,m-2}^{s-1}$; the binary length
is $n=2^{m}$; the number of codewords is $2^k$, where $k= \dd
\sum_{i=0}^r\binom{m}{i}$; the minimum distance is $2^{m-r}$ and
$\RM_s(r-1,m) \subset \RM_s(r,m)$.
\end{theo}

\begin{IEEEproof}
The type  ($4N; \gamma_{r,m}^s, \delta_{r,m}^s$) of the code
$\RM_s(r,m)$  and the minimum distance can be computed from
Proposition~\ref{prop:BQPlotConst}.

To compute the size note that
\begin{eqnarray}
2^k=|\RM_s(r,m)| & = & |\RM_{s-1}(r,m-2)|\times|\RM_{s-1}'(r-1,m-2)|\nonumber\\
                 &   & \times|\hat{\RM}_{s-1}(r-1,m-2)|\times  |\RM_{s-1}(r-2,m-2)|,\nonumber
\end{eqnarray}
\noindent where $\RM_{s-1}'(r-1,m-2)$ and $\hat{\RM}_{s-1}(r-1,m-2)$
are the quaternary linear codes generated by
$\mathcal{G}_{s-1}'(r-1,m-2)$ and
$\hat{\mathcal{G}}_{s-1}(r-1,m-2)$, respectively. Hence,
$$
|\RM_{s-1}'(r-1,m-2)|\times|\hat{\RM}_{s-1}(r-1,m-2)|=2^{2\gamma_{r-1,m-2}^{s-1}+4\delta_{r-1,m-2}^{s-1}}=2^{2k_{r-1}}
$$
So, $k=k_r+2k_{r-1}+k_{r-2}$. Finally, we obtain
\begin{eqnarray}
k & = & \sum_{i=0}^r{m-2 \choose i}+2\sum_{i=0}^{r-1}{m-2 \choose
        i}+\sum_{i=0}^{r-2}{m-2 \choose i} \nonumber\\
  & = & \sum_{i=0}^r{m-1 \choose i}+\sum_{i=0}^{r-1}{m-1 \choose i}\nonumber =
        \sum_{i=0}^r{m \choose i}. \nonumber
\end{eqnarray}

To prove that $\RM_s(r-1,m) \subset \RM_s(r,m)$ notice that from
Lemma~\ref{lemm:PBQ-subset} and since the codes of
Table~\ref{tab:taulaARM(r,1)} and Table~\ref{tab:taulaARM(r,2)}
fulfill the four conditions of this lemma we can conclude by
induction that the code generated by the matrix
$\mathcal{G}_s(r-1,m)$ is a subcode of the code generated by the
matrix $\mathcal{G}_s(r,m)$.
\end{IEEEproof}

For every $0< s\leq \lfloor\frac{m-1}{2}\rfloor $ the  family of
codes  $\RM_s(r,m)$ constructed using the above theorem is
incomplete in the sense that the codes $\RM_s(-1,m)$, $\RM_s(0,m)$,
$\RM_s(m-1,m)$, $\RM_s(m,m)$  do not come from the construction. To
be coherent with all the notations, for $r=-1$, the code
$\RM_{s}(-1,m)$  is defined as the all zero codeword code. For
$r=0$, the code $\RM_{s}(0,m)$  is defined as the repetition code
with only one non zero codeword (the all twos quaternary vector).
For $r=m-1$ and $r=m$, the codes $\RM_{s}(m-1,m)$ and $\RM_{s}(m,m)$
are defined as the even weight code and the whole space
$\Z_4^{2^{m-1}}$, respectively. The construction of the families of
Reed-Muller codes in Theorem~\ref{theo:BQ-plotkin} is based on the
generator matrices and so, for each index $s$, we need a generator
matrix for the codes $\RM_s(-1,m)$, $\RM_s(0,m)$, $\RM_s(m-1,m)$,
$\RM_s(m,m)$.

We will use the following generator matrices: $\mathcal{G}_s(-1,m)=\left(\begin{array}{c}0\cdots 0\\
\end{array}\right)$, $\mathcal{G}_s(0,m)=\left(\begin{array}{c}2\cdots 2\\
\end{array}\right)$, $\mathcal{G}_s(m,m)=I_{2^{m-1}}$.

The generator matrix $\mathcal{G}_s(m-1,m)$ will be recursively
obtained by using the BQ-Plotkin construction
$\mathcal{BQ}(\RM_{s-1}(m-2,m-2),\RM_{s-1}(m-2,m-2),\RM_{s-1}(m-3,m-2))$
(see Definition\ref{defi:BQPlot}).


\begin{prop}
For $m\geq 3$, the matrix $\mathcal{G}_s(m-1,m)$ of
Definition~\ref{defi:BQPlot} associated to
$\mathcal{BQ}(\RM_{s-1}(m-2,m-2),\RM_{s-1}(m-2,m-2),\RM_{s-1}(m-3,m-2))$
is a generator matrix of $\RM_s(m-1,m)$.
\end{prop}
\begin{IEEEproof}
All the rows in  matrix $\mathcal{G}_s(m-1,m)$ are vectors of even
weight. So, to prove that this matrix generates $\RM_s(m-1,m)$ we
only need to check if the dimension is the adequate.

We will prove, by induction on $m\geq 1$, that $\gamma_{m-1,m}^s=1$
and $\delta_{m-1,m}^s=2^{m-1}-1$. The claim is trivially true for
$m=1$ and $m=2$ using the matrices defined in section~\ref{sec:ARM2}
and in (\ref{eq:m=2}). Suppose the claim is true for $m\geq 2$.

 Since the matrix $\mathcal{G}_s(m-1,m)$  is upper triangular,
we must to add the respective values $\gamma$ and $\delta$ of
$\mathcal{G}_{s-1}(m-2,m-2)$, $\mathcal{G}'_{s-1}(m-2,m-2)$,
$\hat{\mathcal{G}}_{s-1}(m-2,m-2)$ and $\mathcal{G}_{s-1}(m-3,m-2)$.

By induction, $\gamma_{m-3,m-2}^{s-1}=1$ and
$\delta_{m-3,m-2}^{s-1}=2^{m-3}-1$. Since
$\mathcal{G}'_{s-1}(m-2,m-2)
=\hat{\mathcal{G}}_{s-1}(m-2,m-2)=\mathcal{G}_{s-1}(m-2,m-2)$ we
have $\gamma_{m-2,m-2}^{s-1}=0$ and
$\delta_{m-2,m-2}^{s-1}=2^{m-3}$.

Hence, we obtain $\gamma_{m-1,m}^s=0+0+0+1=1$,
$\delta_{m-1,m}^s=2^{m-3}+2^{m-3}+2^{m-3}+2^{m-3}-1=2^{m-1}-1$.
\end{IEEEproof}

Note that with these definitions of $\RM_s(-1,m)$, $\RM_s(0,m)$,
$\RM_s(m-1,m)$, $\RM_s(m,m)$, the family of codes $\RM_s(r,m)$,
$0\leq r \leq m$, $0<s \leq \lfloor\frac{m-1}{2}\rfloor$,  fulfills
the four conditions of Lemma~\ref{lemm:PBQ-subset}.

\bigskip

Using Theorems~\ref{theo:plotkin} and \ref{theo:BQ-plotkin} we can construct the $\RM$ codes in the  two
rows of Table~\ref{tab:taulaARM(r,3)}. We do not write the generator
matrices for codes $\RM_0(r,3)$ because they can be directly
obtained from the respective codes for $m = 2$ by using the Plotkin
construction. For the codes in the family $\RM_1(r, 3)$ we present
the generator matrices as a direct application of
Theorem~\ref{theo:BQ-plotkin}:
 \EQ\label{eq:rm3}
\RM_1(0,3):\left(\begin{array}{|cccc}
        2&2&2&2\\
        \hline\end{array}\right);\,\,\RM_1(1,3):\left(\begin{array}{|cccc}\hline
1&1&1&1\\
0&1&2&3\end{array}\right);\,\,\RM_1(2,3):\left(\begin{array}{|cccc}0&0&0&2\\\hline 1&1&1&1\\
0&1&2&3\\
0&0&1&1\\\end{array}\right)
 \EN

\noindent and the remaining code $\RM_1(3,3)$ in the family is the
whole space $\Z_4^{2^{2}}$.

All these codes, after the Gray map, give binary codes with the
same parameters as the codes  $RM(r, 3)$ and with the same
properties described in Theorem~\ref{theo:reed-muller}. In the
case $m = 3$, like in the case $m = 2$ not only these codes have
the same parameters, but they have the same codewords. This is not in this way for all the other values $m > 3$.

Now, from Table~\ref{tab:taulaARM(r,3)} and by using the Plotkin
construction we can construct the two families of the codes
$\RM_s(r,4)$ for $s=0,1$, as it is shown in
Table~\ref{tab:taulaARM(r,4)}. Note that the family of codes
$\RM_1(r,4)$ also can be obtained using the BQ-Plotkin construction
from the family of codes $\RM_0(r,2)$ in
Table~\ref{tab:taulaARM(r,2)}.

\begin{table}\caption{$\RM(r,m)$ codes for $m=4$}\label{tab:taulaARM(r,4)}\begin{center}
$\begin{array}{r||ccccc||c}\cline{2-6} &\multicolumn{5}{c||}{(r,m)}
\\\cline{2-6} & (0,4)&(1,4)&(2,4)&(3,4)&(4,4)
\\\cline{2-6} N & \multicolumn{5}{c||}{(\gamma,\delta)}
\\ \hline\hline
8&(1,0)&(3,1)&(3,4)&(1,7)&(0,8)& \RM_0(r,4)\\
8&(1,0)&(1,2)&(1,5)&(1,7)&(0,8)& \RM_1(r,4)\\\cline{2-6}
\end{array}$\end{center}\end{table}

From~the codes in Table~\ref{tab:taulaARM(r,4)} applying the
Plotkin construction we can construct the two families of
$\RM_s(r,5)$, $s=0,1$, as it is shown in
Table~\ref{tab:taulaARM(r,5)}. The third family in
Table~\ref{tab:taulaARM(r,5)}, $\RM_2(r,5)$, is obtained applying the
BQ-Plotkin construction to the $\RM_1(r,3)$ family of
Table~\ref{tab:taulaARM(r,3)}.

\begin{table}\caption{$\RM(r,m)$ codes for $m=5$}\label{tab:taulaARM(r,5)}\begin{center}
$\begin{array}{r||cccccc||c}\cline{2-7} &\multicolumn{6}{c|}{(r,m)}
\\\cline{2-7} & (0,5)&(1,5)&(2,5)&(3,5)&(4,5)&(5,5)
\\\cline{2-7} N & \multicolumn{6}{c|}{(\gamma,\delta)}
\\ \hline\hline
16 &(1,0)&(4,1)&(6,5)&(4,11)&(1,15)&(0,16) & \RM_0(r,5)\\
16 &(1,0)&(2,2)&(2,7)&(2,12)&(1,15)&(0,16) & \RM_1(r,5)\\
16 &(1,0)&(0,3)&(2,7)&(0,13)& (1,15)&(0,16) &
\RM_2(r,5)\\\cline{2-7}
\end{array}$\end{center}\end{table}

Note that $\RM_0(r,5)$ only can be obtained applying the Plotkin
construction, $\RM_2(r,5)$ only can be obtained applying the BQ-Plotkin
construction, but $\RM_1(r,5)$ can be obtained by using the Plotkin or the
BQ-Plotkin construction.

In general, for $m>1$, the code $\RM_0(r,m)$  can be only obtained
applying the Plotkin construction. For $m$ even and $m$ odd, but
$s\neq \frac{m-1}{2}$, families of $\RM_s(r,m)$ can be obtained
applying the Plotkin or the BQ-Plotkin construction. For $m$ odd and $s=
\frac{m-1}{2}$, $\RM_s(r,m)$ only can be obtained applying the
BQ-Plotkin construction. A question arises at this point, how many
families of Reed-Muller codes can be obtained combining the Plotkin
and the BQ-Plotkin constructions? Next proposition proves that no new
codes appear when we combine both these constructions.

 Given three quaternary linear codes $\mathcal{A}$,
$\mathcal{B}$ and $\mathcal{C}$, we remind that $\pcab$ is the
quaternary linear code obtained applying the Plotkin construction
(see Definition~\ref{defi:PlotDefi}) and $\bqabc$ is the
quaternary linear code obtained by using the BQ-Plotkin
construction (see Definition~\ref{defi:BQPlot}). The following
proposition shows that the two constructions commute.

\begin{prop}
Given four quaternary linear codes $\mathcal{A}$, $\mathcal{B}$,
$\mathcal{C}$ and $\mathcal{D}$, then the codes\\
$\pc(\bqabc),\bqbcd)$ and
$\bq(\pcab,\pc(\mathcal{B},\mathcal{C}),\pc(\mathcal{C},\mathcal{D}))$
are permutationally equivalent.
\end{prop}

The proof is straightforward.

Notice that the same result is true changing the BQ-Plotkin
construction by the quaternary Plotkin construction or the double
Plotkin construction.

From~now on, when we talk about the family of Reed-Muller codes
$\{\RM_s(r,m)\}$ constructed by using the Plotkin and the BQ-Plotkin
constructions we will assume that for $m$ even and $m$ odd, but
$s\neq \frac{m-1}{2}$, these families of codes are obtained applying
the Plotkin construction. For $m$ odd and $s=\frac{m-1}{2}$, the family
of codes is obtained applying the BQ-Plotkin construction.

The following lemma computes the values for the parameters $\gamma$
and $\delta$ of the $\RM_s(r,m)$ codes in the specific case when $m$
is odd, $m\geq 3$ and $s=\frac{m-1}{2}$.

\begin{lemm}\label{lemm:even-code}
For odd $m$, $m\geq 3$ and  $s=\frac{m-1}{2}$ we have the following
values for the parameters $\gamma^s_{r,m}$ and $\delta^s_{r,m}$ of
the  $\RM_s(r,m)$ code built by using the BQ-Plotkin construction
with the generator matrix
 (\ref{matrix_even}):
\begin{description}
    \item[(i)] For odd $r$ it is true that $\gamma^s_{r,m}=0$.
    \item[(ii)] For even $r$ we have $\gamma^s_{r,m}={(m-1)/2 \choose r/2}$.
    \item[(iii)] The following equalities $\delta^s_{m,m}=2^{m-1}$, $\delta^s_{m-1,m}=2^{m-1}-1$ and    $\delta^s_{m-2,m}=2^{m-1}-\frac{m+1}{2}$ are true.
\end{description}
\end{lemm}

\begin{IEEEproof}
Note that by Proposition~\ref{prop:BQPlotConst} it is true that
$\gamma^s_{r,m}=\gamma^{s-1}_{r,m-2}+\gamma^{s-1}_{r-2,m-2}$ with
$\gamma^0_{0,1}=1$ and $\gamma^0_{1,1}=0$. Using induction we can
prove \emph{(i)} and \emph{(ii)}.

Clearly, $\delta^s_{m,m}=2^{m-1}$, $\delta^s_{m-1,m}=2^{m-1}-1$. The
value of $m$ is odd, hence $m-2$ is also odd and
$\gamma^s_{m-2,m}=0$. So, $|\RM_s(m-2,m)|=2^{2\delta^s_{m-2,m}}$
but, also, $|\RM_s(m-2,m)|=2^{2^m-{m\choose m-1} - {m\choose m}}$.
Finally, $2\delta^s_{m-2,m}=2^m-m-1$ and
$\delta^s_{m-2,m}=2^{m-1}-\frac{m+1}{2}$.
\end{IEEEproof}

\bigskip

As it is proved in Theorems~\ref{theo:plotkin} and \ref{theo:BQ-plotkin}
 the constructed families of $\RM$ codes satisfy the same
properties we stated for binary linear Reed-Muller codes in
Theorem~\ref{theo:reed-muller} except for the duality. In the
following Section we will discuss this topic.

Notice that, after the Gray map, the constructed $\RM$ families of quaternary linear
 Reed-Muller codes have not only the same parameters as the
 usual binary linear family of $RM$ codes, but also the
characteristic properties of codes $\RM_s(1,m)$ and $\RM_s(m-2,m)$ as it
 is stated in the following proposition.

\begin{prop}\label{prop:hadamard}
For any integer  $m\geq 1$ and $0\leq s\leq
\lfloor\frac{m-1}{2}\rfloor$, after the Gray map the code
$\RM_s(1,m)$ is a $\Z_4$-linear Hadamard code and the code
$\RM_s(m-2,m)$   is an $\Z_4$-linear extended perfect code.
\end{prop}

\begin{IEEEproof}
From~Theorem~\ref{theo:BQ-plotkin} we have that the codes
$\RM_s(1,m)$, where $0\leq s\leq \lfloor\frac{m-1}{2}\rfloor,$ are
quaternary linear and, under the Gray map, have the parameters of
Hadamard codes. Analogously all the  codes $\RM_s(m-2,m)$, $0\leq
s\leq \lfloor\frac{m-1}{2}\rfloor,$ after the Gray map, are
$\Z_4$-linear and have the parameters of extended perfect binary
codes. By the Krotov classification \cite{Kro00,Kro01}, these
codes could be only $\Z_4$-linear Hadamard and extended
$\Z_4$-linear perfect codes, respectively.
\end{IEEEproof}

\section{Duality}\label{sec:sec4}
For the usual binary linear $RM$ codes we know that the codes $RM(r,m)$ and
$RM(m-r-1,m)$ are dual to each other. The families of $\RM$ codes
have the same property if we use the Kronecker inner product to
define the $\Z_4$-duality.

Throughout this section the notion of duality will be related to the Kronecker inner product defined in~(\ref{kip}).

We begin by studying the duality properties for the family of $\RM$ codes obtained by using the BQ-Plotkin
 construction, that is, we are going to prove the duality relationships for the family
of codes $\RM_s(r,m)$, $m$ odd, $s=\frac{m-1}{2}$, constructed from the family $\RM_{s-1}(r,m-2)$.

Basically, we will prove this fact by induction
but, previously,  we need two technical lemmas. We will use
$\mathcal{G}_s(r,m)$ to refer to the generator matrix of code
$\RM_s(r,m)$; the matrices $\mathcal{G}'_s(r,m)$,
$\mathcal{\hat{G}}_s(m-r-1,m)$ will have the meaning that we
introduced in Definition~\ref{defi:BQPlot} and
$\mathcal{\RM}'_s(r,m)$, $\mathcal{\hat{\RM}}_s(m-r-1,m)$ will be
the codes generated by $\mathcal{G}'_s(r,m)$ and
$\mathcal{\hat{G}}_s(m-r-1,m)$, respectively.

\begin{lemm}\label{lemm:Kip}
Let $\vu,\vv\in \Z_4^N$ be any two vectors such that
$\vu=(\vu_1|\vu_2)$ and $\vv=(\vv_1|\vv_2)$, where
$\vu_1,\vu_2,\vv_1,\vv_2\in \Z_4^{N/2}$. Then,
 \EQ \label{eq:Kipeq}
  \langle (\vu_1|\vu_2),(\vv_1|\vv_2)\rangle_{\otimes N} = \langle \vu_1,\vv_1\rangle_{\otimes
  N/2}+3 \langle \vu_2,\vv_2\rangle_{\otimes  N/2}.
 \EN
\end{lemm}

\begin{IEEEproof}
Straightforward from the Kronecker inner product definition.
\end{IEEEproof}

\begin{lemm}\label{lemm:BQ-preduality1}
Let $m$ be an odd integer, $m\geq 3$, $N=2^{m-1}$ and
$s=\frac{m-1}{2}$. Let $\{\RM_s(r,m)\}$ be the family of $\RM$ codes
obtained in Theorem~\ref{theo:BQ-plotkin} by using the BQ-Plotkin
construction. Then, for each $0\leq r \leq m$, for all $\vu\in
\mathcal{G}'_s(r,m) \backslash \mathcal{\hat{G}}_s(r,m)$ and $\vv\in
\mathcal{\hat{RM}}_s(m-r-1,m)$ we have $\langle
\vu,\vv\rangle_{\otimes N}=0$.
\end{lemm}

\begin{IEEEproof}
We proceed by induction on $m$ beginning with $m=3$. Using
(\ref{eq:rm3}) it is easy to see that the assertion is true for
$m=3$.

For the case when $r$ is odd the statement is trivially true, since
there is nothing to proof. Indeed, from~Lemma~\ref{lemm:even-code},
$\gamma_{r,m}^s=0$ and so $\mathcal{G}'_{s}(r,m)\backslash
\mathcal{\hat{G}}_{s}(r,m) = \emptyset$. Hence, along this proof we
can take $r$ as an even integer.

Now, for $m>3$ and $0< r \leq m-2$, assume by induction hypothesis
that for all $\vx\in \mathcal{G}'_{s-1}(r,m-2) \backslash
\mathcal{\hat{G}}_{s-1}(r,m-2)$ and $\vy\in
\mathcal{\hat{RM}}_s(m-r-3,m-2)$ is $\langle \vx,\vy\rangle_{\otimes
N/4}=0$.

Let $\vv\in \mathcal{\hat{RM}}_s(m-r-1,m)$ and  $\vu\in
\mathcal{G}'_s(r,m)\backslash \mathcal{\hat{G}}_{s}(r,m)$, $0<r\leq
m-2$. We will prove by induction that $\langle
\vu,\vv\rangle_{\otimes N}=0$.

From~Proposition~\ref{prop:BQPlotConst}, we have
$\vu=(\vu_1|\vu_1|\vu_1|\vu_1)+(\zero|\zero|\zero|\vu_4)$, where
$\vu_1\in \mathcal{G}'_{s-1}(r,m-2)\backslash
\mathcal{\hat{G}}_{s-1}(r,m-2)$ and $\vu_4\in
\mathcal{G}'_{s-1}(r-2,m-2)\backslash
\mathcal{\hat{G}}_{s-1}(r-2,m-2)$.

Also, we have $
\vv=(\vv_1|\vv_1|\vv_1|\vv_1)+(\zero|\vv_2|2\vv_2|3\vv_2)+(\zero|\zero|\vv_3|\vv_3)+(\zero|\zero|\zero|\vv_4)
$ with $\vv_1\in \mathcal{\hat{RM}}_{s-1}(m-r-1,m-2)$, $\vv_2\in
\mathcal{RM}'_{s-1}(m-r-2,m-2)$, $\vv_3\in
\mathcal{\hat{RM}}_{s-1}(m-r-2,m-2)$ and $\vv_4\in
\mathcal{\hat{RM}}_{s-1}(m-r-3,m-2)$.

Now, by using Lemma~\ref{lemm:Kip}:
\begin{equation*}
\begin{split}
\kip{\vu}{\vv}{N} & = 8\kip{\vu_1}{\vv_1}{N/4} +
12\kip{\vu_1}{\vv_2}{N/4} +
4\kip{\vu_1}{\vv_3}{N/4} + \kip{\vu_1}{v_4}{N/4} \\
&\quad +  \kip{\vu_4}{\vv_1}{N/4} +  3\kip{\vu_4}{\vv_2}{N/4}
+  \kip{\vu_4}{\vv_3}{N/4} +  \kip{\vu_4}{\vv_4}{N/4}\\
&= \kip{\vu_1}{\vv_4}{N/4} +   \kip{\vu_4}{\vv_1}{N/4}+
\kip{\vu_4}{\vz}{N/4},
\end{split}
\end{equation*}
where $\vz =3\vv_2+\vv_3+\vv_4$.

By induction hypothesis  $\kip{\vu_1}{\vv_4}{N/4}=
\kip{\vu_4}{\vv_1}{N/4}=0$ and so we need only to show that
$\kip{\vu_4}{\vz}{N/4}=0$.

From~Lemma~\ref{lemm:PBQ-subset} we have
$\mathcal{\hat{RM}}_{s-1}(m-r-3,m-2) \subset
\mathcal{\hat{RM}}_{s-1}(m-r-2,m-2) \subset
\mathcal{\hat{RM}}_{s-1}(m-r-1,m-2)$ and so $\vv_3+\vv_4 \in
\mathcal{\hat{RM}}_{s-1}(m-r-1,m-2)$.

As we said at the beginning of the proof, $r$ is even. Therefore, we
have that $m-r-2$ is odd and from~Lemma~\ref{lemm:even-code}, we
obtain $\gamma_{m-r-2,m-2}^{s-1}=0$. Hence,
$\mathcal{G}'_{s-1}(m-r-2,m-2) = \mathcal{\hat{G}}_{s-1}(m-r-2,m-2)$
and $\vz = 3\vv_2+\vv_3+\vv_4 \in
\mathcal{\hat{RM}}_{s-1}(m-r-1,m-2)$. But $\vu_4\in
\mathcal{G}'_{s-1}(r-2,m-2)\backslash
\mathcal{\hat{G}}_{s-1}(r-2,m-2)$. Then, by induction hypothesis,
$\kip{\vu_4}{\vz}{N/4}=0$.

Finally, we prove the statement for $r=0$; $r=m-1$ and $r=m.$

For $r=0$, we proceed by induction. Case $m=3$ is trivially true
taking into account~(\ref{eq:rm3}). We have $\vu\in
\mathcal{G}'_s(0,m)\backslash \mathcal{\hat{G}}_{s}(0,m)$ and so
$\vu$ is the all ones vector $\vu=(1,1,\ldots, 1)$. Any vector
$\vv\in \mathcal{\hat{RM}}_s(m-1,m)$ is generated by the rows of
$\mathcal{\hat{G}}_s(m-1,m)$, where $\mathcal{G}_s(m-1,m)$ is the
matrix defined in~(\ref{matrix_even}). Hence,
$\kip{\vu}{\vv}{N}=\kip{\vu_1}{\vv_4}{N/4}=0$, by induction
hypothesis, since $\vu_1\in \mathcal{G}'_{s-1}(0,m-2)\backslash
\mathcal{\hat{G}}_{s-1}(0,m-2)$ and $\vv_4\in
\mathcal{\hat{RM}}_{s-1}(m-3,m-2)$.

For $r=m-1$ we have $\vu=(0,0,\ldots, 0,1)\in
\mathcal{G}'_s(m-1,m)\backslash \mathcal{\hat{G}}_{s}(m-1,m)$ and
$\vv=(0,0,\ldots, 0)\in \mathcal{\hat{G}}_s(0,m)$, therefore
$\kip{\vu}{\vv}{N}=0$.

Finally, for $r=m$ the claim is trivially true, because the set
$\mathcal{G}'_{s}(m,m)\backslash \mathcal{\hat{G}}_{s}(m,m)$ is
empty.

\end{IEEEproof}

\begin{theo}\label{theo:BQ-duality}
Let $m$ be an odd integer, $m\geq 1$, $N=2^{m-1}$ and
$s=\frac{m-1}{2}$ the set $\{\RM_s(r,m)\}$ be the family of $\RM$
codes obtained in Theorem~\ref{theo:BQ-plotkin} by using the
BQ-Plotkin construction. Then, for each $0\leq r \leq m$, the code
$\RM_s(r,m)$ is a quaternary dual of the  code $\RM_s(m-r-1,m)$.
\end{theo}

\begin{IEEEproof}
Since $|\RM_s(r,m)|\cdot|\RM_s(m-r-1,m)|=2^m$, it suffices to prove
that for every $\vu\in \mathcal{G}_s(r,m)$ and $\vv \in
\mathcal{G}_s(m-r-1,m)$ we have $\kip{\vu}{\vv}{N}= 0$. We proceed
by induction on $m$. The claim is trivially true for $m=1$.

Now, for $m>1$ and $0\leq r \leq m$, assume by induction hypothesis
that for all $\vx\in \mathcal{G}_{s-1}(r,m-2)$ and $\vy\in
\mathcal{G}_{s-1}(m-r-3,m-2)$ it is true that $\langle
\vx,\vy\rangle_{\otimes N/4}=0$.

Let  $\vu\in \mathcal{G}_s(r,m)$ and $\vv\in \mathcal{G}_s(m-r-1,m)$
for any $0<r\leq m$.

When $0<r\leq m-2$ we can use the following expressions for $\vu$
and $\vv$:

\noindent$
\vu=(\vu_1|\vu_1\vu_1|\vu_1)+(\zero|\vu_2|2\vu_2|3\vu_2)+(\zero|\zero|\vu_3|\vu_3)+
(\zero|\zero|\zero|\vu_4), $ where $\vu_1\in
\mathcal{G}_{s-1}(r,m-2)$, $\vu_2\in \mathcal{G}'_{s-1}(r-1,m-2)$
$\vu_3\in \mathcal{\hat{G}}_{s-1}(r-1,m-2)$ and $\vu_4\in
\mathcal{G}_{s-1}(r-2,m-2)$;

\noindent$
\vv=(\vv_1|\vv_1|\vv_1|\vv_1)+(\zero|\vv_2|2\vv_2|3\vv_2)+(\zero|\zero|\vv_3|\vv_3)+
(\zero|\zero|\zero|\vv_4), $ where $\vv_1\in
\mathcal{G}_{s-1}(m-r-1,m-2)$, $\vv_2\in
\mathcal{G}'_{s-1}(m-r-2,m-2)$, $\vv_3\in
\mathcal{\hat{G}}_{s-1}(m-r-2,m-2)$ and $\vv_4\in
\mathcal{G}_{s-1}(m-r-3,m-2)$.

Therefore, applying Lemma~\ref{lemm:Kip} we get
\begin{equation}\label{equacio}
\begin{split}
\kip{\vu}{\vv}{N} & = 8\kipp{\vu_1}{\vv_1} + 12\kipp{\vu_1}{\vv_2}+
4\kipp{\vu_1}{\vv_3} + \kipp{\vu_1}{\vv_4} + \\
&\quad 12\kipp{\vu_2}{\vv_1} + 24\kipp{\vu_2}{\vv_2} +
9\kipp{\vu_2}{\vv_3} +
\kipp{\vu_2}{\vv_4} + \\
&\quad 4\kipp{\vu_3}{\vv_1}+9\kipp{\vu_3}{\vv_2} +
4\kipp{\vu_3}{\vv_3} +
\kipp{\vu_3}{\vv_4} + \\
&\quad \kipp{\vu_4}{\vv_1}+ 3\kipp{\vu_4}{\vv_2} + \kipp{\vu_4}{\vv_3} + \kipp{\vu_4}{\vv_4} \\
&= \kipp{\vu_1}{\vv_4} + \kipp{\vu_2}{\vv_3} + \kipp{\vu_2}{\vv_4} +
\kipp{\vu_3}{\vv_2}
+ \kipp{\vu_3}{\vv_4} + \\
&\quad \kipp{\vu_4}{\vv_1}+ 3\kipp{\vu_4}{\vv_2} +
\kipp{\vu_4}{\vv_3} + \kipp{\vu_4}{\vv_4}.
\end{split}
\end{equation}

All the terms in the above equation are zero as can be seen by
induction hypothesis either directly for $\kipp{\vu_1}{\vv_4}$ and
$\kipp{\vu_4}{\vv_1}$; or using Lemma~\ref{lemm:PBQ-subset} for
$\kipp{\vu_2}{\vv_4}$ and $\kipp{\vu_4}{\vv_2}$; or by
Lemma~\ref{lemm:BQ-preduality1} for $\kipp{\vu_2}{\vv_3}$ and
$\kipp{\vu_3}{\vv_2}$; or applying the inclusions
 $\mathcal{\RM}_{s-1}(m-r-3,m-2)
\subset \mathcal{\RM}_{s-1}(m-r-2,m-2) \subset
\mathcal{\RM}_{s-1}(m-r-1,m-2)$ (see Theorem~\ref{theo:BQ-plotkin})
as in  $\kipp{\vu_3}{\vv_4}$, $\kipp{\vu_4}{\vv_3}$ and
$\kipp{\vu_4}{\vv_4}$.

It remains to prove that the statement is true for two cases:
$\vu\in \mathcal{G}_s(m-1,m)$, $\vv\in \mathcal{G}_s(0,m)$ and
$\vu\in \mathcal{G}_s(m,m)$, $\vv\in \mathcal{G}_s(-1,m)$.

In the first case $\vv=(2,2,\ldots,2)$ and $\mathcal{G}_s(m-1,m)$ is
an even code. In the second case $\vv=(0,0,\ldots,0)$. Therefore, in
both these cases the statement is also true.
\end{IEEEproof}

Now we are going to prove the duality relationship for the families
of codes obtained by using the Plotkin construction.

\begin{theo}\label{duality}
For any integer  $m\geq 2$, let $\{\RM_s(r,m)\}$ be any  families
of $\RM$ codes obtained in Theorem~\ref{theo:plotkin}  by using
the Plotkin construction. Then, for each $0\leq r< m$, the code
$\RM_s(r,m)$  is the quaternary dual of the code $\RM_s(m-r-1,m)$.
\end{theo}
\begin{IEEEproof}
Since $|\RM_s(r,m)|\cdot|\RM_s(m-r-1,m)|=2^m,$ it suffices to show
that for any $\vu\in \mathcal{G}_s(r,m)$ and $\vv \in
\mathcal{G}_s(m-r-1,m)$ we have $\langle \vu,\vv \rangle_{\otimes N}
= 0$. We proceed by induction on $m$. The claim is trivially true
for $m=2$, see (\ref{eq:m=2}). For even $m$ and any $s\leq \lfloor
\frac{m-1}{2}\rfloor$, all the codes of the family $\RM_s(r,m)$ are
constructed by using the Plotkin construction from the family
$\RM_s(r,m-1)$. The same happens when $m$ is odd and  $s<
\frac{m-1}{2}$. But for $m$ odd and  $s =\frac{m-1}{2}$ the codes of
the family $\RM_s(r,m)$ are constructed by using the BQ-Plotkin
construction from $\RM_{s-1}(r,m-2)$. Hence, the initial case for
the induction proof is not only $m=1$, but any $m$ odd and $s
=\frac{m-1}{2}$. This specific case was proved in Theorem
\ref{theo:BQ-duality}.

Now, suppose the claim is true for the family of codes
$\RM_s(r,m-1)$, $0\leq r< m-1$ and $0\leq s \leq \lfloor
\frac{m-2}{2}\rfloor$. Let $\vu=(\vu_1|\vu_1+\vu_2)$, where $\vu_1
\in \mathcal{G}_s(r,m-1)$, $\vu_2 \in \mathcal{G}_s(r-1,m-1)$ and
$\vv=(\vv_1|\vv_1+\vv_2)$, where $\vv_1 \in
\mathcal{G}_s(m-r-1,m-1)$, $\vv_2 \in \mathcal{G}_s(m-r-2,m-1)$.

From~Lemma~\ref{lemm:Kip} we have: $\kip{\vu}{\vv}{N}
=\kip{\vu_1}{\vv_1}{N/2} +
3\kip{(\vu_1+\vu_2)}{(\vv_1+\vv_2)}{N/2} = \kip{\vu_1}{\vv_1}{N/2}
+ 3\kip{\vu_1}{\vv_1}{N/2} +3 \kip{\vu_1}{\vv_2}{N/2} +
3\kip{\vu_2}{\vv_1}{N/2} + 3\kip{\vu_2}{\vv_2}{N/2}= 3
\kip{\vu_1}{\vv_2}{N/2} + 3\kip{\vu_2}{\vv_1}{N/2} +
3\kip{\vu_2}{\vv_2}{N/2}$.

By induction hypothesis, $\langle \vu_1,\vv_2 \rangle_{\otimes
N/2} = 0$ and $\langle \vu_2,\vv_1 \rangle_{\otimes N/2} = 0$.
Moreover, $\langle \vu_2,\vv_2 \rangle_{\otimes N/2} = 0$, since
$\RM_s(r-1,m-1) \subset \RM_s(r,m-1)$.
\end{IEEEproof}

\bigskip


\bigskip

We summarize the properties of the $\RM$ codes in the following theorem:

\begin{theo}
For $m\geq 1$, the quaternary linear Reed-Muller family of codes
$\{\RM_s(r,m)\}$, $0\leq s\leq \lfloor\frac{m-1}{2}\rfloor$, $0\leq r\leq m$,
has the following properties:
\begin{enumerate}
    \item the binary length equals $n=2^m$, $m\geq 1$;
    \item the minimum distance is $d=2^{m-r}$;
    \item the number of codewords is $2^k$, where $\dd k=\sum_{i=0}^r \binom{m}{i}$;
    \item each code $\RM_s(r-1,m)$ is a subcode of the code $\RM_s(r,m)$, $r>0$. The code $\RM_s(0,m)$ is the repetition code with only one nonzero codeword
        (the all twos vector). The code $\RM_s(m,m)$ is the whole space $\Z_4^{2^{m-1}}$ and $\RM_s(m-1,m)$ is the even code (i.e.
        the code with all the vectors of even weight);
    \item the codes $\RM_s(1,m)$ and $\RM_s(m-2,m)$, under the Gray map, are a $\Z_4$-linear
    Hadamard and  a $\Z_4$-linear extended perfect codes respectively;
    \item the code  $\RM_s(r,m)$ is the dual code  of the code
    $\RM_s(m-1-r,m)$ for $-1\leq r\leq m$.
\end{enumerate}
\end{theo}

In this section we used everywhere the Kronecker inner product to
define the duality relationship. But it is also possible to use
the standard definition of inner product given in~(\ref{eq:inner})
and, in this case, instead of the property 6) into the above
Theorem, we obtain an alternative property 6') that we state as a
new result:

\begin{theo}
For $m\geq 1$ and $0\leq s\leq \lfloor\frac{m-1}{2}\rfloor$, given
a quaternary linear Reed-Muller family $\{\RM_s(r,m)\}$ of codes,
$0\leq r\leq m$, there exists a family of quaternary linear
Reed-Muller codes
 $\{\overline{\RM}_s(r,m)\}$, monomially equivalent to $\{\RM_s(r,m)\}$,
such that the code
 $\overline{\RM}_s(r,m)$ is the dual code (by the standard inner
  product) of $\RM_s(m-1-r,m)$ for $-1\leq r\leq m$.
\end{theo}
\begin{IEEEproof}
We have $\langle \vu,\vv \rangle_{\otimes_N} =
 \vu{\cdot}K_N{\cdot}\vv^t =\langle \vu, \vv{\cdot}K_N\rangle$.
 Hence, we define the code $\overline{\RM}_s(r,m)$ as
the code generated by a matrix $\overline{\mathcal{G}}_s(r,m)$,
where
$\overline{\mathcal{G}}_s(r,m)=\mathcal{G}_s(r,m){\cdot}K_N$. Note
that the code generated by the matrix
$\overline{\mathcal{G}}_s(r,m)$ is monomially equivalent to the
code generated by $\mathcal{G}_s(r,m)$.
\end{IEEEproof}

\section{Conclusion}\label{sec:sec5}

 New constructions based on quaternary linear codes has been proposed such that,
 after doing a Gray map, the obtained $\Z_4$-linear codes fulfill the same properties
 and fundamental characteristics as the binary linear $RM$ codes. Apart from the
 parameters characterizing each code an important property which remains in
 these new presented families is that the first order $\RM$ code is, under the Gray map,
 a $\Z_4$-linear Hadamard code and the $(m-2)$-th order $\RM$ code, after the Gray map,
 is a $\Z_4$-linear extended perfect code, like in the usual binary case. So the families
 of codes obtained in the paper, after the Gray map, contain
 the families of $\Z_4$-linear extended perfect and $\Z_4$-linear Hadamard codes introduced
in \cite{Kro00,Kro01}. Moreover, it is important to note that,
after defining the Kronecker inner product, the codes $\RM(r,m)$
and $\RM(m-r-1,m)$ are dual each other like in the binary linear
case.

There are several questions and subjects related to this work
where would be of great interest to go deeply. The first one is
about the generalization of the constructions of $\RM$ codes to
the case of general additive codes, so the case of additive codes
with $\alpha\not =0$. It is known from~\cite{BoRi99} that there
exist $\add$-linear perfect and Hadamard  codes (which are not
$\Z_4$-linear) and these could be the starting point of the new
families. Other questions of interest are related to uniqueness
(up to equivalence) of the codes in a given family of $\RM$ codes,
weight distribution, etc.

%


\end{document}